\documentclass[aps,prb,superscriptaddress,papersize=a4paper,twocolumn,10pt,showpacs]{revtex4-2}
\usepackage{amsmath,amsfonts,amssymb,amsthm}
\usepackage{mathtext,mathtools,mathrsfs, bm, comment}
\usepackage{amssymb, amsmath, amsfonts}

\usepackage{esint,bm,esvect,dsfont}
\usepackage{graphicx,epsfig,psfrag,xcolor}
\usepackage{array,tabularx,tabulary,booktabs,multirow}
\graphicspath{{figures/}}
\usepackage{hyperref}
\usepackage{verbatim}
\usepackage{float}
\usepackage{color}
\usepackage{lipsum}

\usepackage{xcolor}

\definecolor{darkGreen}{RGB}{40,130,50}

\def\kf{k_{\rm F}}

\def\ve{\varepsilon}

\def\bQ{\mathbf{Q}}

\def\bk{\mathbf{k}}

\def\ve{\varepsilon}

\def\br{\mathbf{r}}

\def\bk{\mathbf{k}}

\def\fv{v_\textrm{F}}

\begin{document}

\title{Density of electronic states, resistivity and superconducting transition temperature in density-wave compounds with imperfect nesting}	
        \author{A.~V.~Tsvetkova}
\affiliation{National University of Science and Technology MISIS, Moscow, 119049 Russia}
	\author{Ya.~I.~Rodionov}
		\affiliation{Institute for Theoretical and  Applied Electrodynamics, Russian Academy of Sciences, Moscow, 125412 Russia}
\affiliation{National Research University Higher School of Economics, Moscow 101000, Russia}

	\author{P.~D.~Grigoriev}
		\affiliation{L.D. Landau Institute of Theoretical Physics, RAS, Chernogolovka, 143423, Russia}
\affiliation{National University of Science and Technology MISIS, Moscow, 119049 Russia}

\date{\today}

\begin{abstract}
We study the effects of imperfect nesting in the density wave (DW) state on various electronic properties within a simple 2D tight-binding model. The discussed model reflects the main features of quasi-1D metals where the DW emerges. We show that a DW with imperfect nesting leads to unusual singularities in the quasiparticle density of states and to a power-law renormalization of the superconducting critical temperature.  Our results are derived at arbitrary large antinesting and may help to understand the phase diagram of the  wide class of density-wave superconductors. We also compute the conductivity tensor in a wide temperature range, including the DW transition, and obtain a satisfactory agreement with the experimental data on rare-earth trichalcogenides and many other DW materials.
\end{abstract}


\maketitle
\section{Introduction}

The interplay between superconductivity (SC) and charge or spin density wave (DW) attracts a vast research activity. The SC-DW competition and coexistence appears in various strongly-correlated electron systems, including the high-temperature cuprate \cite{XRayNatPhys2012, XRayPRL2013,Eduardo2014,XRayCDWPRB2017,Tabis2014,Science2015Nd,Wen2019,Miao2019,Miao2021} and iron-based \cite{Medvedev2009,ReviewFePnictidesAbrahams,ReviewFePnictides2,Wu2015,Sun2016,Han2020} superconductors, NbSe$_2$ \cite{CDWSCNbSe2,NbSe2NatComm,Feng2023Interplay} and other transition-metal and rare-earth
di and poly-chalcogenides \cite{Hamlin2009,Zocco2015,Chikina2020,Zeng2022}, 
various organic superconductors \cite{Ishiguro1998,AndreiLebed2008-04-23,CDWSC,Hc2Pressure,Vuletic,Kang2010,ChaikinPRL2014,Gerasimenko2014,Yonezawa2018}, and many other materials 
(see, e.g., Refs. \cite{Review1Gabovich,ReviewGabovich2002,MonceauAdvPhys,Pouget2024} for reviews). 

The DW in metals creates a spatial modulation of the charge or spin density of conducting electrons with the wave vector $\boldsymbol{Q}$ and opens a gap $\Delta$ at the Fermi level in the electronic spectrum, which reduces the electron energy \cite{Gruner1994}. Depending on the electron dispersion $\epsilon (\boldsymbol{k})$ in this metal, two different DW states are possible. The first, called the perfect nesting of the Fermi surface (FS), occurs when $\epsilon (\boldsymbol{k})+\epsilon (\boldsymbol{k}+\boldsymbol{Q})<\Delta$ for all $\boldsymbol{k}$ on the FS. Then the DW covers the entire FS and the metal becomes a semiconductor in a DW state. In this case a uniform SC on a DW background is hardly possible, but SC may appear via a spatial segregation with DW, as happens in various materials \cite{KresinReview2006,ChaikinPRL2014,Yonezawa2018,Seidov2023,Campi2015}. This SC heterogeneity can be visualized by the scanning tunneling microscopy (STM) and spectroscopy \cite{KresinReview2006,Hoffman2011,InhBISCCO,InhBISCO2009,Massee2009,InhCaFeAs,Cho2019,InhCaFeAs2018,Song2012,InhFeSe,InhCaFeAs2018}, by the local diamagnetic probe \cite{DiaLaSrCuO,DiaYBCOInh}, or detected and analyzed using the resistivity anisotropy measurements \cite{Sinchenko2017,Seidov2018,KesharpuCrystals2021,Kochev2021,Kochev2023,Grigoriev2023FeSe}. The spatial DW-SC phase segregation usually appears on a large length scale, greater than the DW coherence length $\xi_{\rm DW}=\hbar v_F/(\pi\Delta )$, as it happens in organic superconductors \cite{ChaikinPRL2014,Yonezawa2018,Kochev2023}, or on the microscopic scale, i.e. in the form of DW soliton walls \cite{BrazKirovaReview,GG,GrigPhysicaB2009}. 
 
The second scenario of DW-SC coexistence appears in the case of imperfect FS nesting when $\epsilon (\boldsymbol{k})+\epsilon (\boldsymbol{k}+\boldsymbol{Q})>\Delta_0$ for some $\boldsymbol{k}$ at the FS. In this case the metallic conductivity survives in a DW state till $T\to 0$, being anisotropically reduced, e.g., as in various rare-earth three-chalcogenides \cite{Sinchenko2014} and many other materials \cite{Review1Gabovich,ReviewGabovich2002,MonceauAdvPhys,Gruner1994}. One could expect an exponential decrease of the superconducting transition temperature $T_c$  in the presence of the DW background even in the imperfect-nesting case, because according to the BCS theory in the weak-coupling regime
\begin{equation}
	T_c\sim \omega_D \exp (-1/g\nu_F)
	\label{Tc0}
\end{equation}
exponentially depends on the electron density of states (DoS) at the Fermi level $\nu_F$ multiplied by the coupling constant $g$, with the Debye frequency $\omega_D$ in the pre-exponential factor. As the DW opens a gap, at least in some parts of FS, the DoS $\nu_F$ is reduced by the DW, leading to destructive SC-DW interference \cite{Levin1974,Balseiro1979}.
However, usually, one observes a more complicated SC-DW interplay: the superconducting transition temperature $T_c$ is the highest at the quantum critical point (QCP) where the density wave (DW) gets suppressed by some external parameter, such as doping level \cite{XRayNatPhys2012, XRayPRL2013,XRayCDWPRB2017,Tabis2014,Science2015Nd,Wen2019}, pressure \cite{Medvedev2009,ReviewFePnictidesAbrahams,ReviewFePnictides2,Wu2015,Ishiguro1998,AndreiLebed2008-04-23,CDWSC,Hc2Pressure,Vuletic,Kang2010,ChaikinPRL2014}, cooling rate \cite{Gerasimenko2014,Yonezawa2018}, disorder \cite{NbSe2NatComm}, etc. The $T_c$ dome-like shape near the DW QCP results from the enhancement of electron-electron (e-e) interaction $g\to g_*(\boldsymbol{Q})$ in the Cooper channel by the critical DW fluctuations, which can be described as the SC vertex renormalization in the language of Feynman diagram technique \cite{Bychkov1966,Chubukov2013pairing}. Note that this vertex renormalization changes its momentum dependence and may favor the unconventional superconductivity \cite{Tanaka2004,Nickel2005,GGPRB2007}. 

Below we consider only the second scenario of the uniform DW in the case of imperfect nesting. What is stronger, the SC coupling renormalization $g\to g_*(\boldsymbol{Q})$ or the DoS reduction $\nu_F\to \nu_{F*}$ by the DW? The answer to this question  dramatically affects the SC transition temperature in the weak-coupling regime given by Eq. (\ref{Tc0}).

According to the theoretical calculations \cite{Bychkov1966,Tanaka2004,Nickel2005,Chubukov2013pairing} the SC coupling enhancement $g\to g_*(\boldsymbol{Q})$ is not very strong, especially far from the DW QCP, when the DW order parameter $\Delta $ is larger than the SC energy gap. On the contrary, the gapped FS area in the DW state is, usually, rather large, as visualized by ARPES measurememts in various compounds \cite{Schmitt2008,Moore2010,Kang2006,Qian2008,Ryu2018,Hoesch2019,Miao2021,Luo2022,Kato2022,Kato2023,Chikina2023}. How the observed increase of $T_c$ in the presence of the DW, implying the increase of the product $g\nu_F$, is then possible? 

The insight resolving this apparent inconsistency is proposed in Ref. \cite{Grigoriev2008} where the electronic DoS in the DW state with slightly imperfect nesting is calculated in the mean-field approximation. The DoS is shown to become unexpectedly large at the Fermi level, close to the DoS without DW, even if the ungapped FS pockets are very small. This happens due to a strong renormalization of electron spectrum by the DW. Hence, $T_c$  affected by the DW background,  combined with the coupling-constant renormalization $g\to g_*(\boldsymbol{Q})$, can be even higher than  the SC transition temperature $T_{c0}$ without DW. The calculations in Ref. \cite{Grigoriev2008} are limited by the very small ungapped FS pockets and are performed under some approximations about the electron dispersion.

In this paper we generalize analysis of the DoS in the presence of the DW  for the case of arbitrary imperfect nesting, when the size of ungapped FS parts may be large. Although the main goal of our calculations is to estimate the change of SC transition temperature by the DW background, our results are also useful  for other electronic properties of DW compounds, even without superconductivity. For example, the DoS enters the electronic part of specific heat and other thermodynamic and transport electron properties, which can be measured.




\section{Model}
\label{SecModel}

The DW state usually appears in strongly anisotropic quasi-1D metals due to their good FS nesting \cite{Gruner1994,MonceauAdvPhys,Jerome2024}. In many other DW compounds there is a hidden quasi-one-dimensionality, although their electronic transport exhibits a 2D isotropy along the conducting layers.  An example of such a hidden quasi-1D electron spectrum are the rare-earth tritellurides \cite{Brouet2008,Sinchenko2014} or tetratellurides \cite{Pathak2022RTe4,Meng2022}, where the FS consists of two pairs of perpendicular quasi-1D warped sheets originating from the Te $p_x$ and $p_y$ orbitals, giving almost tetragonal symmetry of electronic properties, broken by the CDW \cite{Sinchenko2014}. Of course, there are some DW materials with closed quasi-2D rather than open quasi-1D FS, but even they have some nearly flat FS parts, such as transition metal
dichalcogenides \cite{Inosov2008} or high-Tc cuprates \cite{Miao2019,Miao2021}. 
Below we consider quasi-1D metals as a quite generic system bearing the DW.

In layered quasi-2D metals due to crystal periodicity the
electron dispersion is given by the Fourier series
\begin{gather}
\label{spectrum0}
    \ve(\mathbf{k}) = -2t_x\cos ({ak_x}) - 2t_y \cos ({bk_y}) \\
    - \sum_{i,j\geq 2} 2t_{ij}\cos({ak_{x}i})\cos({bk_{y}}j) \notag ,
\end{gather}
where $a$ and $b$ are the lattice constants in $x$ and $y$ directions,  while $t_x,\ t_y$ and $t_{ij}$ are the hopping amplitudes. The tunneling amplitude along the $z$ axis is considerably smaller than that along the conducting layers $x,y$ and is discarded, as it does not play any role in the analysis below. However, we still assume a 3D layered material where the DW fluctuations are weaker than in true low-dimensional systems, so that the mean-field DW description is valid, at least qualitatively. Usually, the last term in Eq. (\ref{spectrum0}) is much smaller than the first two and is also omitted,  corresponding to the tight-binding model~\cite{abrikosov2017fundamentals,ziman1979principles}. 

In quasi-1D metals, where the $x$-axis represents an easy-conducting chain direction and $t_x\gg t_y$, it is often convenient to linearize the $k_x$ electron dispersion near the Fermi level. Then the free-electron dispersion without
magnetic field can be written as \cite{GorkovFISDW1984,MonceauAdvPhys,Grigoriev2005,Grigoriev2008}
\begin{gather}
\label{spectrum}
    \ve(\mathbf{k}) = \fv(|k_{x}| - k_{F}) - 2t_y \cos ({bk_y}) - 2t_{y}'\cos({2bk_{y}}),
\end{gather}
where the first term represents the electron dispersion along the $x$-direction linearized near the Fermi surface.  
We drop all the other hopping amplitudes in $y$ - direction, as they are small by the parameter $t_y/t_x\ll 1$. 

\begin{figure}[t!]
	\centering
\includegraphics[width=0.9\columnwidth]{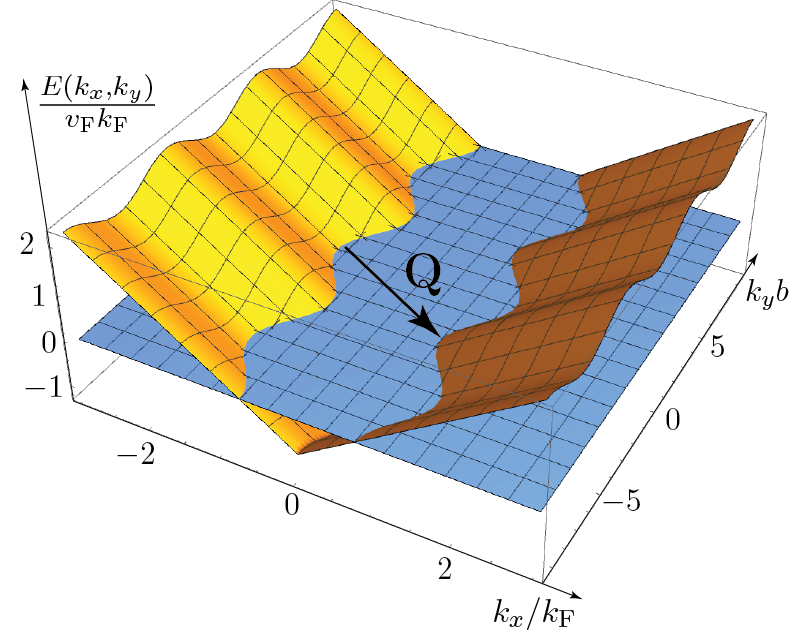}
	\caption{The electron spectrum $E(k_x,\ k_y)$ given by Eq.~\eqref{spectrum} is shown by yellow surface. The FS (two Fermi wavy curves in the $k_x-k_y$ plane) is formed by the intersection of this spectrum with the horizontal blue plane of fixed electron energy equal to the chemical potential or Fermi energy $k_{\rm F}v_{\rm F}$.}
\label{spectr1}
\end{figure}

Without the last \emph{antinesting} $t_y^\prime$ term in Eq. ~\eqref{spectrum} the perfect nesting condition 
\begin{gather}
\label{nesting}
    \ve(\bk)+\ve(\bk+\mathbf{Q}) = 0
\end{gather}
is satisfied at wave vector $\mathbf{Q} = (2\kf, \pi/b)$ (see Fig.~\ref{spectr1}).  The electron spectra and FS near $k_{\rm F}$ and $ -k_{\rm F}$ are nested  with the nesting vector $\mathbf{Q}$. 
The outlined quasi-1D system exhibits the DW formation as the result of a nesting-driven Fermi surface instability. 
 This leads to the opening of a gap $\Delta$ 
 in the quasiparticle spectrum. However, this simple picture 
 undergoes a dramatic change once we take into 
 account the antinesting amplitude $t_y^\prime$.

Although in quasi-1D metals $t_y^\prime \sim t_y^2/t_x \ll t_y$, the retained term $t_y^\prime$ rather than $t_y$ competes with the energy gap $\Delta$ in the resulting ground state equations.
The $2t_y^\prime$ term in~\eqref{spectrum} is called “antinesting” because it violates the nesting condition~\eqref{nesting}. In what follows, we discard the possible influence of $t_y^\prime$ on the nesting vector $\mathbf{Q}$ and discuss the legitimacy of this approach in section \ref{SecDiscuss}.

As seen from Fig.~\ref{spectr1}, the FS consists of two warped unconnected curves separated by vector Q. 
This allows us to split the quasiparticle Fock's space into two disconnected parts corresponding to the neighborhood of each Fermi sheet 
\begin{gather}
\begin{split}
\label{psi}
    \psi(\mathbf{r}) &= \sum\limits_{\bk \in U(\bk_F)} a(\bk)e^{i\bk\br} + \sum\limits_{\bk \in U(-\bk_F)}a(\bk)e^{i\bk\br} \\
    &\equiv
    \sum\limits_{\bk} a_1(\bk)e^{i\bk\br} + \sum\limits_{\bk}a_2(\bk)e^{-i\bk\br}, 
\end{split}
\end{gather}
where $a(\bk)$ are the annihilation operators, and $a_{1,2}(\bk)$ are the same operators with momenta positioned in the respective neighborhoods $U(\pm\bk_F)$.
As usual in Peierls's transition physics, we assume that the deformation of the lattice has a static nature. The lattice deformation potential mixes two different parts (see Eq. \eqref{psi}) of the operators' Fock's space and adds one more term in the Hamiltonian. Accordingly, the electron Hamiltonian can be written as follows
\begin{equation}
\label{ham}
    \hat{H} = \hat{H}_0 + \hat{H}_{\rm int}
\end{equation}
with the free-electron part in momentum representation being
\begin{equation}
\label{ham1}
    \hat{H}_0 = \sum\limits_{\bk} a_1^\dag(\bk) a_1(\bk) \ve(\bk) + \sum\limits_{\bk} a_2^\dag(\bk) a_2(\bk) \ve(\bQ+\bk),
\end{equation}
and the interaction part
given by
\begin{equation}
\label{ham2}
    \hat{H}_{\rm int} = \Delta \sum\limits_{\bk} \left( a_1(\bk)a_2^\dag(\bk) + a_1^\dag(\bk) a_2(\bk) \right),
\end{equation}
where $\Delta$ is the DW order parameter, proportional to the amplitude of lattice deformation for the charge-density wave (CDW). In the physics of Peierls transition $\Delta \equiv \Delta(T)$ becomes the temperature-dependent gap in the quasiparticle spectrum.   

\begin{figure*}[htpb]
	\centering
\includegraphics[width=1.8\columnwidth]{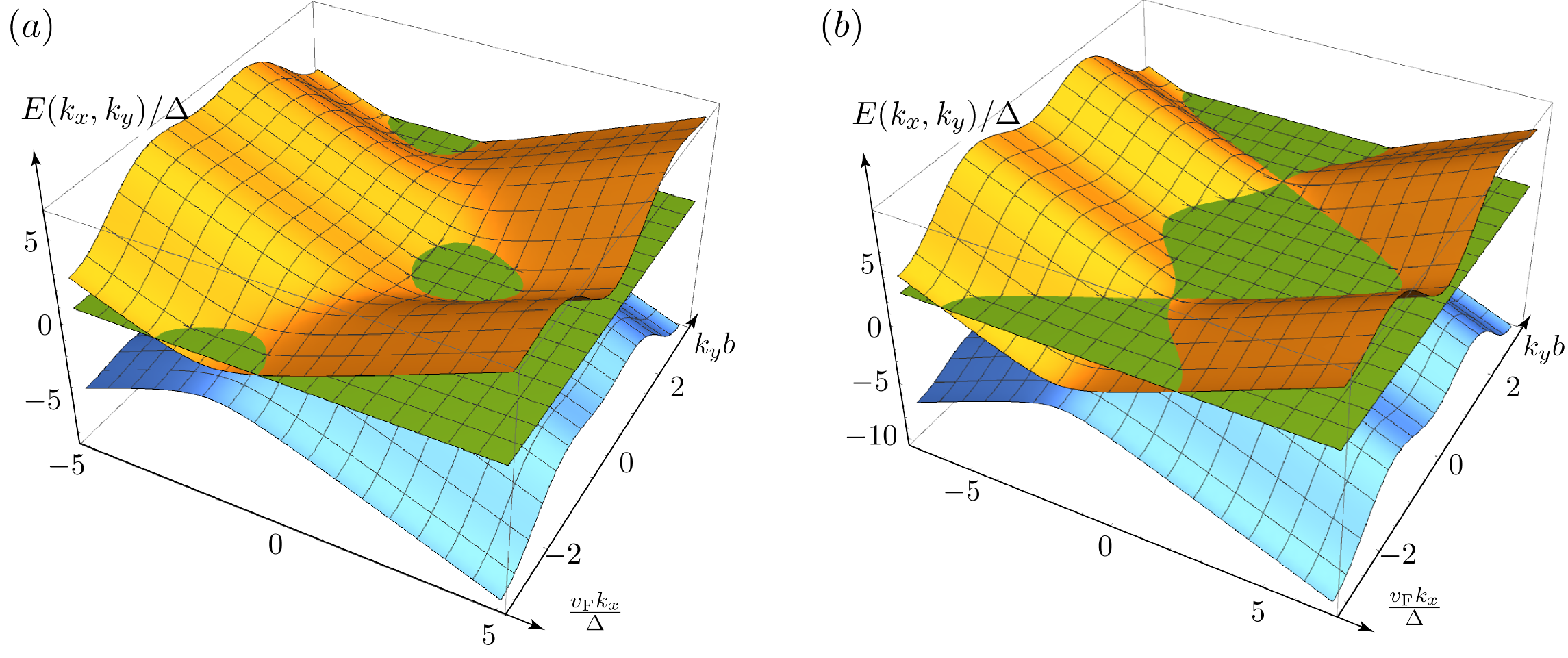}
	\caption{(a) The quasiparticle spectrum (yellow and blue) in the presence of a CDW with $\Delta_1=0.8\Delta$. The horizontal green plane corresponds to the constant energy cross section $\ve = \Delta_1 \equiv 2t_y^\prime$. The closed pockets
     are clearly seen as green puddles above the yellow surface. (b)  The same as in Fig. (a) but at $\Delta_1=1.5\Delta$ and the energy level $\ve = \Delta + \Delta_1$, corresponding to the critical value at which the DoS manifests logarithmic divergence.}
\label{fpockets1}
\end{figure*}

Diagonalization of the Hamiltonian ~\eqref{ham} leads to the standard gapped spectrum
\begin{equation}
\label{spectr2}
    E(\bk) = \frac{\epsilon(\mathbf{k}) + \epsilon(\mathbf{k+Q})}{2} \pm \sqrt{\frac{(\epsilon(\mathbf{k})-\epsilon(\mathbf{k+Q}))^2}{4}+\Delta^2}.
\end{equation}
Substituting Eq. \eqref{spectrum} to Eq. \eqref{spectr2} we obtain the following quasiparticle spectrum in the presence of the CDW 
\begin{equation}
\label{energy}
    E_{\pm}(k_x,k_y) = -2t_{y}'\cos{2k_{y}b} \pm \sqrt{(k_{x}v_{\rm F}-t_{y}\cos{k_{y}b})^2+\Delta^2}.
\end{equation}
We illustrate it in Fig.~\eqref{fpockets1}.  Note that even if all $t_{ij}=0$ in Eq. (\ref{spectrum0}), in Eq. (\ref{energy})
$t_{y}'\sim t_y^2/t_x\neq 0$ and comes from the nonlinearity of electron dispersion along the  $x$-direction.

\section{Quasiparticle DoS in a DW state at arbitrary imperfect nesting}
\label{SecDOS}

\subsection{General formulas}

\begin{figure}[h!]
	\centering
\includegraphics[width=0.9\columnwidth]{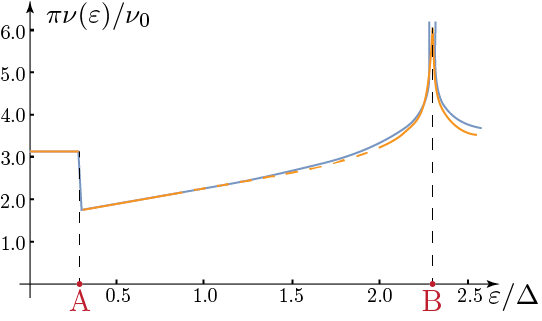}
	\caption{Comparison of the numerical DoS (blue curve) with analytical results (orange curve). The singularities of the function are clearly seen. At point A, $\ve = \Delta_1 - \Delta$ the DoS function has a jump (Eq. \eqref{DOS5}) and at point B $\ve = \Delta_1 + \Delta$, the DoS has a logarithmic divergence (Eq. \eqref{DOS3}). We used the value $\Delta_1 = 1.3 \Delta$ for the  antinesting term. Here $\nu_0=[\pi v_{\rm F} b]^{-1}$ is the DoS in the metallic state without DW.} 
\label{fcomparisonres}
\end{figure}

To understand the behavior of the transport properties of the system we need to compute the single particle density of states (DoS)
\begin{equation}
\label{deltaint}
    \nu(\ve) = \sum_{\mathbf{k},\sigma} \delta(\ve - E_\sigma (\mathbf{k})) \equiv \sum\limits_{\sigma}\int \delta(\ve - E_\sigma(\mathbf{k})) \; \frac{dk_xdk_y}{(2\pi)^2},
\end{equation}
where $\delta(x)$ is the Dirac $\delta$-function. Integration is performed over the  Brillouin zone. Due to the symmetry of gapped spectrum \eqref{spectr2} the integration region can be reduced to the quarter of the Brillouin zone. After the integration over $k_x$ the expression \eqref{deltaint} becomes
\begin{gather}
    \nu(\ve) = \frac{1}{\pi^2 v_{\rm F}}\int\limits_0^{\pi /2}\frac{|\ve+\Delta_1 \cos 2k_y b|}{\sqrt{(\ve+\Delta_1\cos 2k_y b)^2-\Delta^2}} \: dk_y
\label{DOS2}
\end{gather}
with the notation $\Delta_1\equiv 2t_y^\prime $ introduced for convenience. Although the integral~\eqref{DOS2} can only be evaluated numerically (see Fig. \ref{fcomparisonres}), 
one can obtain analytical results in some regions where the DoS reveals intriguing (singular) behavior. These cases are discussed in the next subsection. 

\subsection{The behavior of DoS near singularities}

The DoS is an even function of energy as can be easily checked from expression \eqref{DOS2}. This function has two singularities. The jump at point $\Delta_1 - \Delta$ and the divergence at point $\Delta_1 + \Delta$, corresponding to the appearance and the merger of the closed pockets on the surface of fixed energy (see Fig.~\ref{fpockets1}).

The character of the singularities in the DoS can be understood from the structure of the integrand in Eq.~\eqref{DOS2}.
Namely, we see that the square root in the denominator of the integrand of ~\eqref{DOS2} has branch points at 
$\ve = \pm\Delta\pm\Delta_1$ where electron pockets of constant energy  appear or merge, as in Fig. \ref{fpockets1}(b).
The corresponding analytical expression
for the DoS  near the threshold energy $\Delta+\Delta_1$ reads:
\begin{gather}
\label{DOS3}
\begin{split} 
&\nu(\Delta + \Delta_1 + \delta\ve) = \frac{1}{\pi^2 v_{\rm F}b}\bigg[2\arcsin{\sqrt{\frac{\Delta_1}{\Delta + \Delta_1}}}  \\ 
&+\sqrt{\frac{\Delta}{4\Delta_1}}\left( \ln{\frac{32\Delta}{\Delta_1 + \Delta}} - \ln{\frac{\delta\epsilon}{\Delta_1}} \right)
\bigg].
\end{split}
\end{gather}
As we see from Eq. ~\eqref{DOS3}, the DoS has a logarithmic singularity at the threshold energy $\ve = \Delta +\Delta_1$, given by the sum of the DW gap $\Delta$ and of the antinesting parameter $\Delta_1$. This corrects and generalizes the qualitative analysis presented in Ref.
 \cite{Grigoriev2008} (Fig. 3), where the log-singularity was predicted at $\ve = \Delta$ instead of $\Delta +\Delta_1$.

There exists another threshold $\ve= \Delta_1-\Delta$. We need to address two possible situations: $\Delta< \Delta_1$ and
$\Delta>\Delta_1$. We obtain the following results (see Appendix A):
\begin{subequations}
\begin{align}
       &\nu(\Delta_1-\Delta+\delta\ve) =\notag\\
        &\frac{1}{\pi v_F b}
        \Big[\big[1-f(\delta\ve)\big]+f(\delta\ve)\theta(-\delta\ve)\Big],
       \quad \Delta<\Delta_1 \label{DOS5}\\
            &\nu(\Delta-\Delta_1+\delta\ve) = \frac{1}{\pi  v_F b}f(-\delta\ve)\theta(\delta\ve),\quad
    \Delta>\Delta_1 \label{DOS51}\\
     &f(\delta\ve) = \frac{1}{2}\sqrt{\frac{\Delta}{\Delta_1}}-\frac{\delta\ve}{16}\sqrt{\frac{\Delta}{\Delta_1}}\frac{3\Delta_1+\Delta}{\Delta_1}.\notag
\end{align}
\end{subequations}
The resulting graph, juxtaposing the analytical formulas \eqref{DOS3}-\eqref{DOS51} in the vicinity of singularities and numerical computation of~\eqref{DOS2} is presented in Fig.~\ref{fcomparisonres}.


\section{Superconducting transition temperature in the presence of DW}
\label{SecTc}

Now we analyze how the discussed above singular behavior of the DoS and the antinesting term affect the superconducting transition temperature $T_c$  in the presence of the DW background.

The equation for $T_c$ is given by the BCS integral
\cite{abrikosov2017fundamentals} 
\begin{equation}
\label{Tcint}
     \frac{2}{g} = 2 \int^{\omega_D}_{0} \frac{d\xi}{\xi}\tanh{\frac{\xi}{2T_c}}\nu(\xi),
 \end{equation}
where $g$ is the electron-phonon coupling constant in the BCS theory and $\omega_D$ is the cutoff at Debye frequency. The Eq. \eqref{Tcint} reproduces the corresponding integral derived from the linearized Gor'kov equations for superconductivity  in the presence of the DW background \cite{GGPRB2007,Grigoriev2008}).

We perform the calculation at Debye frequency $\omega_D< \Delta_1+\Delta$, i.e. 
the integration domain doesn't include the log-singularity of the DoS. This can be the case in  e.g., rare earth tritellurides \cite{banerjee2013charge,Brouet2008}, where $\omega_D\sim 0.015$\ eV$\sim 170$K and the CDW energy gap $\Delta \approx 0.3$eV. The superconductive gap is assumed to vanish, since we position ourselves just above the SC transition temperature.  In this case the integration~\eqref{Tcint} can be performed analytically with the help of expression~\eqref{DOS5}. We obtain the following formula for $T_c$:
\begin{gather}
    \nonumber
     T_c = \left(\frac{\Delta_1 - \Delta}{\omega_D}\right)^{\frac{\delta\nu}{\nu_0}} T_0,
     \quad 0< \Delta_1-\Delta<\omega_D<\Delta+\Delta_1;\\
     T_0 = \omega_D e^{-\frac{1}{\nu_0 g}}, \; \nu_0 = \frac{1}{\pi  v_F b} ,\; \delta\nu = \frac{1}{\pi  v_F b} \sqrt{\frac{\Delta}{\Delta_1}}. 
 \label{Tc}
 \end{gather}
Here $\nu_0$ is the electron DoS without DW in quasi-1D metals  and $\delta\nu$ is the DoS jump at $\ve = \Delta_1-\Delta$.

\begin{figure}[h]
	\centering
\includegraphics[width=0.9\columnwidth]{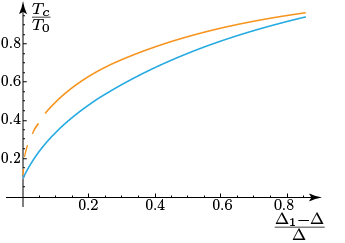}
	\caption{Comparison of the numerical $T_c$ (blue curve) with analytical result~\eqref{Tc} (orange curve), where $\Delta_1$ is the antinesting term and $\Delta$ is the Peierls energy gap. The dashed part of the analytical curve corresponds to the region $\Delta_1 -\Delta \lesssim T_c$ where the analytical relation~\eqref{Tc} doesn't work.}
\label{fcomparisonTc}
\end{figure}

As expected, we reproduced the unmodified BCS nonanalytic exponential DoS-dependence of $T_c$. However, we also obtained an interesting change in the pre-exponential part. 
This is precisely where the antinesting term in the initial Hamiltonian~\eqref{spectrum} comes on stage. The comparison of the analytical result~\eqref{Tc} with the numerical calculation of integral~\eqref{Tcint} is presented in Fig. ~\ref{fcomparisonTc}.
Formulae~\eqref{Tc} and the plot in Fig.~\ref{fcomparisonTc} are the main physical results of this section.

The result~\eqref{Tc} is interesting from the following perspective. As the substantial part of FS area is squeezed under the DW gap, one would intuitively expect an additional exponential suppression \cite{Levin1974,Balseiro1979} of $T_c$ by a DW, due to the DoS $\nu$ entering the exponential decrement of $T_c$ in Eq.~\ref{Tc}. Surprisingly, this does not happen. We see the non-trivial renormalization of the pre-exponential factor instead. This result is important for the explanation of a rather high superconducting transition temperature $T_c$ in numerous DW superconductors (see, e.g., \cite{Review1Gabovich,ReviewGabovich2002,MonceauAdvPhys,Denholme2017}), including high-Tc cuprates. If the suppression of $T_c$ by the DW was exponential, the enhancement of e-e interaction in the Cooper channel by the DW fluctuations \cite{Bychkov1966,Chubukov2013pairing,Tanaka2004,Nickel2005} would not compensate for it, and one would always observe a decrease of $T_c$ in the presence of the DW, contrary to the most experiments showing a $T_c$ dome-like shape \cite{Review1Gabovich,ReviewGabovich2002,MonceauAdvPhys,XRayNatPhys2012, XRayPRL2013}.

\section{Temperature dependence of resistivity}
\label{SecR}

In this section we analyze how the obtained in Sec. \ref{SecDOS}  nontrivial behavior of DoS manifests itself  in the temperature dependence of the electric  conductivity $\sigma (T)$. The single particle DoS depends on temperature via the DW gap $\Delta (T)$ and affects the resistivity $R(T)$ via the electron mean free time already in the Born approximation. Often such an influence is completely discarded, e.g., in Ref. ~\cite{Sinchenko2014} where the DoS of the compound was assumed constant in the transport analysis. Our study is motivated by numerous experiments on resistivity $R(T)$ in DW metals with imperfect nesting \cite{Gruner1994,Gabovich2001,MonceauAdvPhys,Sinchenko2014,Xu2024}.

To compute the diagonal components of the conductivity tensor, we employ the Kubo-Greenwood formalism. We
start from Kubo - formula, based on quasiparticles of DW state described within the mean-field approximation
\begin{gather}
\label{conduct}
\begin{split}
    &\sigma_{ij} = - e^2\int \frac{d\ve}{2\pi} \frac{\partial f(\ve)}{\partial \ve} \\
    &\times\sum_{\sigma = \pm}\int \frac{d^2 \bk}{(2\pi)^2} v_{i\sigma}(\bk) v_{j\sigma}(\bk) \langle G^A_\sigma(\ve,\bk) G^R_\sigma(\ve,\bk)\rangle,
\end{split}
\end{gather} 
where $f(\ve)$ is the Fermi distribution function, $v_{i\sigma}(\bk) \equiv \nabla_{k_i}E_\sigma(\bk)$ is $i$th component of the velocity vector operator (see Eq.\ref{energy}), $G^{R,A}_\sigma(\ve, \bk) = [\ve-E_\sigma(\bk)\pm i0]^{-1}$ are the single particle retarded and advanced Green's functions respectively and $e$ is the electron's charge. 
The angular brackets denote the disorder averaging. Below we discard the difference between scattering time and transport scattering time. 
This is valid for the short-range disorder, including short-wavelength phonons, screened Coulomb interaction and usual short-range crystal defects. In terms of disorder averaging, this corresponds to the substitute $\langle G^RG^A\rangle\rightarrow \langle G^R\rangle\langle G^A\rangle$, where the disorder-averaged Green's function is given by the relation:
$\langle G_\sigma^{R,A}(\ve,\bk)\rangle = [\ve-E_\sigma(\bk)\pm i(2\tau)^{-1}]$, and $\tau$ is the elastic scattering time.

The elastic scattering rate $(2\tau)^{-1}$ of electrons by impurities and by other short-range crystal defects is given by the imaginary part of the quasiparticle Green's function self-energy, which we take in the first Born approximation. For simplicity, we assume the $\delta$-correlated disorder potential $u(\br)$,  diagonal in quasiparticle basis:
\begin{subequations}
    \begin{align}
    & \langle u(\br)u(\br^\prime)\rangle = g^2n_{\rm imp}\delta(\br-\br^\prime),\\
    &\frac{1}{2\tau(\ve)} = \frac{\pi}{2} g^2 n_{\rm imp}\!\! \int \delta\big(\ve- E_\sigma(\bk) \big) \frac{d^2\bk}{(2\pi)^2} = \frac{\pi}{2} n_{\rm imp} g^2 \nu (\ve ).
\label{time1}
\end{align}
\end{subequations}
Here, $n_{\rm imp}$ is the concentration of short-range crystal defects, mainly impurities, and $g$ is the disorder potential amplitude.
Integrating~\eqref{conduct} over energy  $\ve$ yields the following expressions for the diagonal terms of conductivity tensor (see also Appendix C):
\begin{subequations}
 \begin{align}
 \label{condTxx}
     \sigma_{xx}(T) &= \frac{ v_{\rm F}^2}{2T} \sum\limits_\sigma\int \tau[E_\sigma(\bk)]\zeta_\sigma (\bk, T) \frac{d\bk}{(2\pi)^2} \\
     \label{condTyy}
     \sigma_{yy}(T) &= \frac{ b^2}{2T} \sum\limits_{\sigma}\int \tau[E_\sigma(\bk)][2t_y \sin{b k_y}]^2 \zeta_\sigma(\bk, T)\frac{d\bk}{(2\pi)^2},\\
      \zeta_\sigma (\bk, T) &= \frac{e^2}{4\cosh^2{\frac{E_\sigma(\bk,T)}{2T}}} \frac{(v_{\rm F} k_x - t_y \cos{bk_y})^2}{(v_{\rm F} k_x - t_y \cos{bk_y})^2 + \Delta^2(T)}\notag.
\end{align}
 \end{subequations}
The temperature dependence of the Peierls gap $\Delta(T)$ is taken in the mean-field approximation and Landau theory of the second-order phase transition as
\begin{gather}
     \Delta(T) = \Delta_0 \sqrt{1 - \frac{T}{T_{\rm DW}}},
 \end{gather}
where $T_{\rm DW}$ is the DW phase transition temperature.


Apart from impurity scattering, there is an additional contribution to the electron scattering rate from electron-phonon interaction. The full scattering rate is the algebraic sum of the impurity and phonon scattering.
The electron-phonon scattering rate is described by the Bloch-Gruneisen law \cite{ziman1979principles}:
 \begin{gather}
 \begin{split}
     \frac{1}{\tau_{\rm ph}(T)} = \frac{1}{\tau_{\rm BG}} \cdot F\left(\frac{\theta_D}{T} \right), \\
     F(z) = \int^z_0 \frac{z^5 \, dz}{\left(e^z - 1 \right) \left( 1 - e^{-z} \right)}.
\label{time2}
 \end{split}
 \end{gather}
Here, the Bloch-Gruineisen time $\tau_{\rm BG}$ contains the sound velocity, the ion mass, and Debye temperature $\theta_D$ of the compound.
Formula~\eqref{time2} takes into account the effective number of phonons on which the carriers scatter. The full scattering rate is given by the sum of the elastic scattering rate  by crystal disorder and phonon rate:
\begin{gather}
   \tau^{-1} =\tau_0^{-1}+\tau_{\rm ph}^{-1}
\label{tauFull}
\end{gather}

\begin{figure}[h]
	\centering
\includegraphics[width=1\columnwidth]{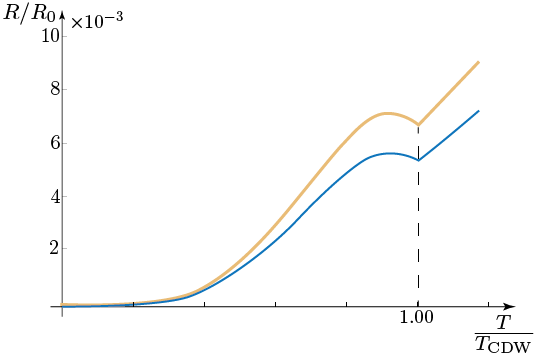}
	\caption{Dimensionless components of the anisotropic resistivity: $10\times R_{xx}$ (blue curve) and $R_{yy} \times t_y^2b^2/v_{\rm F}^2$ (orange curve) as a functions of temperature extracted from Eq.~\ref{condTxx},~\ref{condTyy}. Parameter $\tau^{-1}_{\rm BG}$ entering the scattering rate~\eqref{time2} is taken such that the residual resistance ratio parameter for $R_{xx}$ is $\approx 70$; $R_0 = b/(e^2v_{\rm F}\tau_{\rm BG})$.} 
\label{conductT}
\end{figure}
Taking into account Eq.~\ref{tauFull}, we numerically integrate
Eqs.\ref{condTxx},\ref{condTyy}. The resultant resistivity components are presented in Fig. \eqref{conductT}. To perform calculations, we took realistic DW parameters \cite{Sinchenko2014} $\Delta_0 = 0.27$ eV, $T_c = 0.03$  eV, $t_y = 0.37$ eV, $t'_y =  0.16$ eV. 
Our results presented by plots in Fig.~\ref{conductT} demonstrate a qualitative consistency with the numerous experimental data obtained in DW metals \cite{Gruner1994,MonceauAdvPhys,Gabovich2001} (e.g., see Fig. 1 of Ref. \cite{Sinchenko2014} or Fig. 3.12 of the monograph \cite{Gruner1994}). Note that the previous calculations, shown in Fig. 4 of Ref. \cite{Sinchenko2014}, predicted a much stronger increase of resistivity $R(T)$ below the CDW phase transition $T_{\rm DW}$.  At first glance, one indeed expects a much larger resistivity increment below $T_{\rm DW}$ than shown in  Fig.~\ref{conductT} as large FS parts disappear due to a DW gap in the quasiparticle spectrum, given by Eq. \eqref{spectr2}. Here we resolve this inconsistency by taking into account the $T$-dependence of the electron mean-free time $\tau$, which also strongly increases below $T_{\rm DW}$ even in the Born approximation. The latter takes place due to the decrease of electron DoS at the Fermi level caused by the DW. The calculation of the temperature dependence of resistivity in a DW state with imperfect nesting is the second main result of our paper.

\section{Discussion}
\label{SecDiscuss}

We studied the influence of DW with arbitrarily imperfect nesting on the SC critical temperature and explored how it manifests itself in the temperature dependence of resistivity in the framework of a general model of a realistic quasi-1D metal.   
We have discovered that the DW with imperfect nesting does not affect the (DoS dependent) exponential increment in the  BCS formula~\eqref{Tc0} for $T_c$. Although such a DW strongly reduces the FS by creating an energy gap covering part of its area, it renormalizes the preexponential factor in the SC transition temperature only.
The discussed renormalization comes from the change in the electron dispersion and in the FS geometry due to the antinesting term $t_y^\prime$ in the electron spectrum ~\eqref{spectrum}.
Finally, the temperature dependence of the conductivity tensor is analyzed in terms of the discussed model. We show that the nontrivial DoS leads to qualitative agreement of our results with the experimental data in rare-earth tritellurides \cite{Sinchenko2014} and in many other DW compounds \cite{Gruner1994,MonceauAdvPhys}.  

\begin{figure}[ht]
	\centering
\includegraphics[width=0.9\columnwidth]{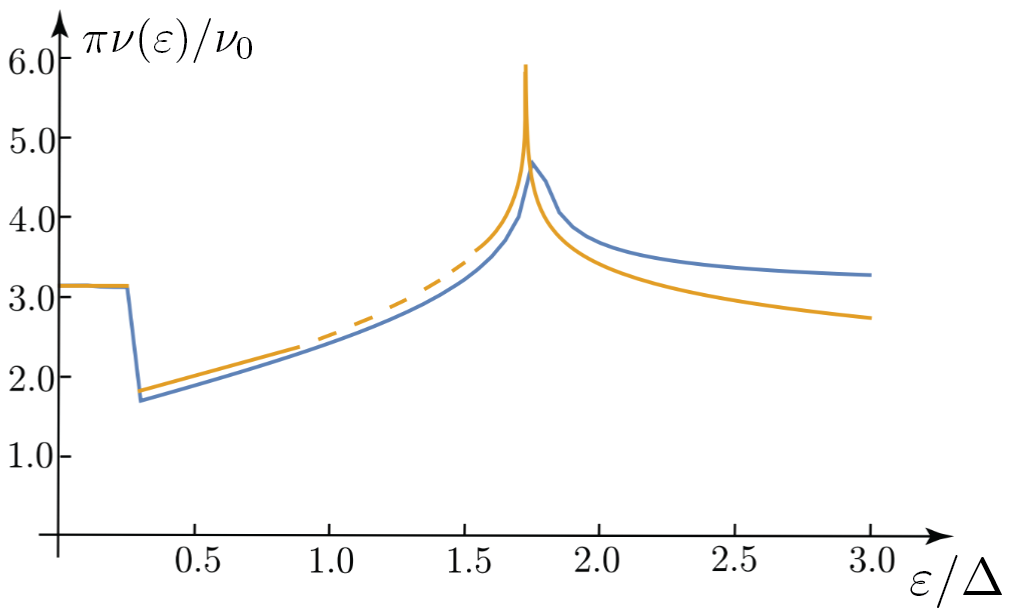}
	\caption{Comparison of the numerical DoS (blue curve)  with the displacement vector $\delta\mathbf{Q}$ taken into account and the analytical curve following from Eqs.~\ref{DOS5}-~\ref{DOS51} (orange line)  taken at the same physical parameters. The general shape of the DoS function is not  significantly affected (also see Fig. \ref{fcomparisonres} for comparison).}
\label{fcorrectvector}
\end{figure}

We calculated the quasiparticle DoS at arbitrary ratio of the antinesting parameter $2t_b^{\prime}\equiv \Delta_1$ in electron dispersion \eqref{spectrum} to the DW energy gap $\Delta$. The obtained DoS $\nu (\ve) $ has two singularities: a jump down at $\ve =\Delta_1-\Delta $ and a logarithmic divergence at $\ve =\Delta_1+\Delta $, which corrects and generalizes the previous studies \cite{Grigoriev2008}. The obtained DoS differs considerably from that in the BCS theory, which describes well the DoS of the CDW metals with ideal nesting \cite{Gruner1994}. In the BCS theory the first singularity is absent, while the second singularity is a square-root divergence of DoS at $\ve =\Delta $, which appears in Eq. \eqref{DOS2} only at $\Delta_1=0$. For imperfect nesting this square-root divergence shifts by $\Delta_1=2t_b^{\prime}$ and weakens to a logarithmic one due to the extra integration over the interchain direction $y$. However, this divergence does not disappear completely, as the antinesting cosine term $2t_b^{\prime}\cos (2k_yb)$ results in van Hove singularities at its boundaries $\pm 2t_b^{\prime}$ after the integration over $k_y$ in Eq. \eqref{DOS2}.

Next, the following important remark is in order.
We completely discarded the influence of the antinesting term $t_y^\prime$ on the DW vector $\mathbf{Q}$ throughout the paper. However, the latter may shift the DW vector from  its initial value $\mathbf{Q} = (2k_F, \pi/b)$ at $t_y^\prime \neq 0$ \cite{Guster2020,Pouget2024}.  
Inevitably, the question arises if the initial nesting vector $\mathbf{Q}$ can be used for our study of the influence of the DW formation on the electron DoS and physical observables, such as $T_c$ and $R(T)$. 
In the absence of ideal nesting, the DW wave vector $\mathbf{Q}$ maximizes the Kubo susceptibility 
\begin{gather}
    \chi(T,\mathbf{Q})=
    -\int\frac{d\bk}{(2\pi)^2}\frac{f[\ve(\bk-\mathbf{Q}]-f[\ve(\bk)]}{\epsilon(\bk-\mathbf{Q})-\epsilon(\bk)},
\end{gather}
since the condition of the Peierls transition corresponds to the equation 
$\chi(\mathbf{Q})=g^{-1}$, where $g$ is the electron-phonon coupling constant.
Taking a realistic temperature $T = 100 K$ and a considerable antinesting $\Delta_1=1.3\Delta$, we computed the displacement $\delta \mathbf{Q}$ of the DW wave vector. Our calculations give
$\delta Q_x/Q_x \approx -0.02$ and $\delta Q_y/Q_y \approx -0.2$ in a reasonable agreement with earlier calculations \cite{Guster2020,Pouget2024}. 
The corresponding numerical plot of the DoS, where the  calculated displacement vector $\mathbf{Q}$ is taken into account, is presented in Fig. ~\ref{fcorrectvector} (blue curve). For comparison, we show the analytical curve  given by Eqs. ~\ref{DOS5}-~\ref{DOS51} (the shift $\delta\mathbf{Q}$ is discarded), similar to that in Fig.3, in the same plot.
As we see from Fig.~\ref{fcorrectvector}, the positions of singularities and the quantitative shape of both curves are not significantly affected by the change of nesting vector $\delta\mathbf{Q}$ discussed above. 
One concludes that for a not too large antinesting term, the change of the DoS due to the shift of nesting vector $\mathbf{Q}$ is negligible. Hence, although the shift of the DW wave vector from its value $\mathbf{Q} = (2k_F, \pi/b)$ at ideal nesting is considerable, it barely affects the quasiparticle DoS and the observable quantities such as superconducting $T_c$ and the temperature dependence of resistivity.

 Our formula \eqref{Tc} for the SC transition temperature $T_c$ in the presence of the DW differs from Eq. 47 of Ref. \cite{Grigoriev2008}. It also predicts the non-exponential $T_c$ decrease (power law) as a function of $\Delta_1-\Delta $ (see  Fig. \ref{fcomparisonTc}). Unlike predicted in  Ref. \cite{Grigoriev2008} (Eq. 47), this power differs from $1/2$. This difference comes from two sources. First, our formula \eqref{Tc} is not restricted by the limiting case $|\Delta_1-\Delta |\ll \Delta$, as in Ref. \cite{Grigoriev2008}. Second, it considers a different interval of the DW gap $\Delta$ and antinesting term $2t_b^{\prime}\equiv \Delta_1$ with respect to the Debye frequency $\omega_D$, which serves as an upper limit of the integral in Eq. \eqref{Tcint}. Ref. \cite{Grigoriev2008} however, addresses the opposite limiting case $\omega_D\gg \Delta_1+\Delta $, which is relevant to organic superconductors \cite{AndreiLebed2008-04-23}, where the DW transition temperature is rather low ($T_{\rm DW} \sim 10$K, so that $\Delta_{\rm DW} \lesssim 50$K). A different situation arises in cuprate high-Tc superconductors \cite{XRayNatPhys2012, XRayPRL2013,Eduardo2014,XRayCDWPRB2017,Tabis2014,Science2015Nd,Wen2019,Miao2019,Miao2021}, rare-earth tritellurides \cite{Hamlin2009,Zocco2015},  nickel- and iron-based high-Tc superconductors, iron-based \cite{Guo2023,Medvedev2009,ReviewFePnictidesAbrahams,ReviewFePnictides2,Wu2015,Sun2016,Han2020},  NbSe$_2$\cite{CDWSCNbSe2,NbSe2NatComm,Feng2023Interplay} and other transition-metal and rare-earth
di and poly-chalcogenides \cite{Hamlin2009,Zocco2015,Chikina2020,Zeng2022}, where the DW transition temperature $T_{\rm DW}\gtrsim 100$K, and the DW gap $\Delta \gtrsim 0.1eV \gtrsim \hbar \omega_D$ even in the SC phase. Therefore, the regime $\Delta_1-\Delta<\omega_D<\Delta+\Delta_1$ in which the analytical results of in Sec. \ref{SecTc} are obtained is valid for wide class of compounds. 

Eq. \eqref{Tc} can be rewritten in a slightly different form:
\begin{equation}
      \frac{T_c}{T_0} = \left(\frac{\Delta_1}{\omega_D} \right)^{\alpha } \left[1-\frac{\Delta}{\Delta_1} \right]^{\alpha } ,\quad  \alpha\equiv\sqrt{\frac{\Delta}{\Delta_1}}. 
 \label{Tc1}
 \end{equation}
The obtained power law in Eq. \eqref{Tc1} contains the exponent $\alpha \equiv \sqrt{\Delta /\Delta_1}$, which is always less than unity. However, in the limit $\Delta_1-\Delta\ll \Delta$ this exponent $\alpha \to 1$ is considerably larger than $\alpha = 1/2$ presented in Eq. (47) of Ref. \cite{Grigoriev2008}. The prefactor $\left(\frac{\Delta_1}{\omega_D} \right)^{\alpha } \sim 1$ and does not change $T_c$ strongly. Hence, in the limit $\Delta_1-\Delta\ll \Delta$ Eq, \eqref{Tc1} predicts a lower $T_c$ than Eq. (47) of Ref. \cite{Grigoriev2008}. The physical reason for this stronger $T_c$ decrease is as follows.
The limit $\omega_D \gg \Delta_1+\Delta $ addressed in \cite{Grigoriev2008} implies that the integration domain in \eqref{Tcint} for $T_c$
spans the region
$\xi > \Delta_1+\Delta$
where the DoS  has a logarithmic singularity and satisfies the condition $\nu >\nu_0$.
In our calculations in Sec. \ref{SecTc} and in Eqs. \eqref{Tc} and \eqref{Tc1}, derived at $\omega_D<\Delta+\Delta_1$, this interval does not contribute to the integral \eqref{Tcint}, lowering the resulting  $T_c$. Hence, at $\omega_D > \Delta_1+\Delta $ the SC transition temperature is even higher than in Eqs. \eqref{Tc} and \eqref{Tc1}.

For large antinesting $\Delta_1 > 4\Delta$ the exponent $\alpha < 1/2$ (see Eqs. \eqref{Tc} and \eqref{Tc1}). 
Combined with the additional factor $\left(\Delta_1/\omega_D \right)^{\alpha } \gtrsim 1$ in Eq. \eqref{Tc1}, this makes the predicted $T_c$ even larger than in Ref. \cite{Grigoriev2008}, but still lower than $T_0$. 
Finally, at $\Delta_1 \gg \Delta$ the predicted ratio $T_c/T_0 \to 1$, so that the $T_c$ renormalization caused by the reduction of electron DoS at the Fermi level by DW gap is negligible.
Therefore, due to a known effect of additional enhancement of SC coupling coming from DW fluctuations \cite{Bychkov1966,Tanaka2004,Nickel2005,Chubukov2013pairing}, we predict the  increase of SC transition temperature by DW, in agreement with numerous experiments on SC -- DW interplay in the high-Tc superconductors and in many other materials.

Lastly, we should emphasize the following counterintuitive result: the smallness of the resistivity increment at and below the DW transition temperature $T_{\rm DW}$ (see Fig. \ref{conductT}). This result agrees quite well with experimental observations in various materials (see e.g., Figs. 1 and 2 of Ref. \cite{Ru2008}, Fig. 1 of Ref. \cite{Sinchenko2014}, Fig. 3.12 of Ref. \cite{Gruner1994}). Naively, one may expect a rather large resistivity increment at $T_{\rm DW}$ in these materials, because only tiny FS parts remain ungapped, as observed by ARPES (e.g., see Fig. 3a of Ref. \cite{Schmitt2011} for TbTe$_3$). However, as we  show in Sec. \ref{SecR}, the resistivity increment at $T_{\rm DW}$ is much smaller than one could  expect from the dramatic reduction of the FS. The physical explanation is as follows. Despite the fact that the FS shrinks, the reduction of quasiparticle density at the Fermi level affects both: the number of charge carriers and  their mean free time. When computing the conductivity, these two effects happen to compensate for each other in the first approximation. The small resistivity increment at $T_{\rm DW}$ still appears (see Fig. \ref{conductT}) due to the renormalization of quasiparticle spectrum, given by Eq. \eqref{spectr2}, which reduces the mean electron velocity at the Fermi level. This, in turn, reduces the conductivity,  proportional to the mean square of electron velocity at the Fermi level \cite{abrikosov2017fundamentals,ziman1979principles} (see also Eq. \ref{conduct} above). To our knowledge, this simple qualitative picture was missing before. For example, in Ref. \cite{Sinchenko2014} an alternative explanation of the  unexpectedly small resistivity increment at $T_{\rm DW}$ was proposed. It is based on a very slow grow (possibly, due to CDW fluctuations) of the CDW order parameter $\Delta$ as temperature decreases just below $T_{\rm DW}$. However, not only this very slow increase of $\Delta$ contradicts  the Landau theory of phase transitions  but also the experimental observations of the temperature dependence of elastic X-ray scattering intensity at the CDW wave vector (see Figs. 6 and 7 of Ref. \cite{Ru2008}). On the contrary, our theoretical description of the temperature dependence of resistivity agrees well with all available experimental data in these compounds.  

\medskip
To summarize, we have studied the electronic properties of the general model of a density wave compound with imperfect nesting for arbitrary value of the antinesting parameter. We have computed its DoS and explored its influence on such important observable characteristics as superconducting $T_c$ and the temperature dependence of resistivity. Our results help to shed some light on the phase diagram and transport properties of various density-wave superconductors, including the high-Tc cuprates, transition-metal and rare-earth polychalcogenides, nickel- and iron-based high-Tc superconductors, and other promising materials.

 
\acknowledgements{A.V.T. acknowledges the Foundation for the Advancement
of Theoretical Physics and Mathematics “Basis” for Grant No. 22-1-1-24-1 and the NUST “MISIS” Grant No. K2-2022-025 in the framework of
the federal academic leadership program Priority 2030. P.D.G.
acknowledges the State Assignment No. FFWR-2024-0015. }

\appendix

\section{DoS with imperfect nesting}
We put $v_{\rm F} = 1 ,\; b = 1$ in what follows and restore correct units in the final formulae.
In all subsequent calculations, the integration region is a quarter of the Brillouin zone due to the symmetry of the gap spectrum \eqref{spectr2}.
Let us first assume that the antinesting term is not small and satisfies the condition: 
\begin{equation}
    2t_{y}^{'} = \Delta_1 > \Delta.
\end{equation}
One can analytically compute the DoS in the vicinity of $\Delta + \Delta_1$: $\epsilon = \Delta + \Delta_1 + \delta\epsilon\; , \; \delta\epsilon \ll \Delta$. After integrating over $k_x$ the integral \eqref{deltaint} becomes:
\begin{equation}
   2 \int_{0}^{\pi} \frac{\Delta + \delta\epsilon + \Delta_1(1+\cos{2k_y})}{\sqrt{(\Delta + \delta\epsilon + \Delta_1(1+\cos{2k_y}))^2 - \Delta^2}} \, \frac{dk_y}{(2\pi)^2}.
\end{equation}
Subtracting singular terms, we obtain: 
\begin{gather}
\begin{split}
 &2 \int_{0}^{\pi} \left[ \frac{b + 2\cos^2{k_y}}{\sqrt{(b + 2\cos^2{k_y})^2 - b^2}} - \sqrt{\frac{b}{4}} \frac{1}{|\cos{k_y}|} \right] \,\frac{dk_y}{(2\pi)^2} \\
 & + 2 \sqrt{\frac{b}{2}} \int_{0}^{\pi} \frac{1}{\sqrt{2\cos^2{k_y} + a}} \, \frac{dk_y}{(2\pi)^2},\\
& b = \frac{\Delta}{\Delta_1}, \; a = \frac{\delta\epsilon}{\Delta_1}.
\end{split} 
\label{uglyintA}
\end{gather}
The first part of integral \eqref{uglyintA} can be evaluated easily with the substitution $t = \cos{k_y},\; dk_y = \frac{dt}{\sqrt{1 - t^2}}$: 
\begin{equation}
\begin{split}
\label{Ans1B}
    &2 \int_{0}^{\pi} \left[ \frac{b + 2\cos^2{k_y}}{\sqrt{(b + 2\cos^2{k_y})^2 - b^2}} - \sqrt{\frac{b}{4}} \frac{1}{|\cos{k_y}|} \right] \, \frac{dk_y}{(2\pi)^2} =\\  
    &\int^1_0 \left[ 
    \frac{2t}{\sqrt{b+t^2}} + \frac{b - \sqrt{b^2 + bt^2}}{t\sqrt{b+t^2}} \right]\, \frac{dt}{(2\pi)^2\sqrt{1-t^2}} =\\ 
    &\frac{1}{(2\pi)^2}\int^1_0 \frac{dt}{t\sqrt{1-t^2}} 
    \left[ \frac{b + 2t^2}{\sqrt{b+t^2}} - \sqrt{b} \right] =\\ 
    &\frac{1}{2\pi^2} \left[\arcsin{\frac{1}{\sqrt{1+b}}} - \frac{\sqrt{b}}{2}\ln{\left(1 + \frac{1}{b}\right)}
    \right].
\end{split}
\end{equation}
The second part of the integral \eqref{uglyintA} reads 
\begin{gather}
\begin{split}
    &\frac{2}{(2\pi)^2}\sqrt{\frac{b}{2}} \int_{0}^{\frac{\pi}{2}}
    \frac{ds}{\sqrt{2\sin^2{s}+a}} =\\ 
    &\frac{2}{(2\pi)^2}\sqrt{\frac{b}{2}} \left[ \int_{0}^{\frac{\pi}{2}} \left( \frac{ds}{\sqrt{2}\sin{s}} - \frac{ds}{\sqrt{2}s}
    \right) +  \int_{0}^{\frac{\pi}{2}} \frac{ds}{\sqrt{2s^2 + a}} \right] =\\ 
    &\frac{\sqrt{b}}{(2\pi)^2}\left (
    \ln{4\sqrt{2}} - \ln{\sqrt{a}}
    \right) \\ 
    &s = k_y - \frac{\pi}{2}, \; \frac{s}{k_y} \ll 1.
\end{split} 
\label{Ans2B}
\end{gather}



Summing up \eqref{Ans1B} and \eqref{Ans2B}, we obtain the final form of the DoS given by Eq. ~\eqref{DOS3} in the main body of the paper.



Now we put $\ve = \Delta_{1} - \Delta + \delta\ve ,\; \Delta_1 > \Delta , \; \delta\ve \ll \Delta$. There are two different cases. When $\delta\ve  > 0$ or $\delta\ve = 0$ the pockets of the FS from the lower branch have not yet appeared. The integral \eqref{DOS2} reads
\begin{gather}
\nonumber \int_{0}^{\arccos{\sqrt{b+\frac{a}{2}}}} \frac{a - b + 2\cos^2{k_y}}{\sqrt{(a - b + 2\cos^2{k_y})^2 - b^2}} \, \frac{dk_y}{(2\pi)^2}, \\
    b = \frac{\Delta}{\Delta_1}, \; a = \frac{\delta\ve}{\Delta_1}.
\end{gather}
To compute the integral \eqref{DOS2} at $\delta\ve = 0$, we use the substitution $t = \cos{k_y}$:
\begin{gather}
\begin{split}
    &\int_{0}^{\arccos{\sqrt{b}}} \frac{- b + 2\cos^2{k_y}}{\sqrt{(- b + 2\cos^2{k_y})^2 - b^2}} \, \frac{dk_y}{(2\pi)^2} = \\ 
    &\int_{0}^{\arccos{\sqrt{b}}} \frac{2\cos^2{k_y} - b}{2\sqrt{(\cos^2{k_y} - b)(\cos^2k{k_y})}} \, \frac{dk_y}{(2\pi)^2} =\\ 
    &\int_{\sqrt{b}}^{1} \frac{2t^2 - b}{2t\sqrt{(t^2 - b)}\sqrt{1 - t^2}} \, \frac{dt}{(2\pi)^2} = 
    \frac{1}{8\pi}\left(1 - \frac{\sqrt{b}}{2}\right).
\end{split}
\label{res1B}   
\end{gather}
When $\delta\ve < 0$, two small pockets appear in the vicinity of the point $q_y =\pi/2$. Consequently, they contribute to the DoS. The corresponding contribution looks as follows:
\begin{gather}
\begin{split}
   &\int_{\frac{\pi}{2} - \arccos{\frac{\sqrt{a}}{2}}}^{\frac{\pi}{2}} \frac{a - b + 2\cos^2{k_y}}{\sqrt{(a - 2b + 2\cos^2{k_y})(a + 2\cos^2{k_y})}} \, \frac{dk_y}{(2\pi)^2} =\\ 
    &\int^{\arcsin{\frac{\sqrt{|a|}}{2}}}_{0} \frac{-|a| - b + 2\sin^2{s}}{\sqrt{(-|a| - 2b + 2\sin^2{s})(-|a| + 2\sin^2{s})}} \, \frac{ds}{(2\pi)^2}, \\
    &s = k_y - \frac{\pi}{2}, \; \frac{s}{k_y} \ll 1.
\end{split}
\end{gather}
Now, we expand $\sin s$ and $\arcsin{\frac{\sqrt{|a|}}{2}}$ over a small parameter $a$:
\begin{gather}
     \nonumber
     \int_{0}^{\sqrt{\frac{|a|}{2}}} \frac{-|a| - b + 2s^2}{\sqrt{(-|a| - 2b + 2s^2)(-|a| + 2s^2)}} \, \frac{ds}{(2\pi)^2} =
    \\ \int_{0}^{\sqrt{\frac{|a|}{2}}} \frac{-b}{\sqrt{2b(|a| - 2s^2)}} \, \frac{ds}{(2\pi)^2} = -\sqrt{b}\frac{1}{16\pi}.
\label{res2B}
\end{gather}
Finally, combining Eq. \eqref{res1B} and Eq. \eqref{res2B} we obtain Eq.~\eqref{DOS5} for the density of states.

Now let us calculate the DoS with the energy
$\ve = \Delta - \Delta_1 + \delta\ve, \; \Delta > \Delta_1.$
The integral for DoS \eqref{DOS2} reads
\begin{gather}
   \nonumber
   \int^{\pi/2}_0 \frac{a + \frac{b}{2} - \sin^2{k_y}}{\sqrt{(a + b - \sin^2{k_y})(a - \sin^2{k_y})}}\,\frac{dk_y}{(2\pi)^2} , \\
    a = \frac{\delta\epsilon}{2\Delta_1}, \;
    b = \frac{\Delta}{\Delta_1}, \;
    \sin{k_y} = a\sin{t}.
\end{gather}
The completely analogous calculations yield Eq.~\ref{DOS51}.
\medskip

\section{SC critical temperature}
In the equation for $T_c$ \eqref{Tcint} the r.h.s. can be represented as the sum of simple integrals, where we use the step-like structure of the density of states from Eqs.~\eqref{DOS5},~\eqref{DOS51}: 
 \begin{gather}
\begin{split}
     &\frac{1}{g} = \nu \int^{\Delta_1 - \Delta}_{T_c} \frac{d\xi}{\xi} + (\nu - \delta\nu) \int^{\omega_D}_{\Delta_1 - \Delta} \frac{d\xi}{\xi} = \\ 
     &\nu \ln{\left(\frac{\Delta_1 - \Delta}{T_c}\right)} + (\nu - \delta\nu) \ln{\left(\frac{\omega_D}{\Delta_1 - \Delta}\right)} = \\ &\nu \ln{\frac{\omega_D}{T_c}} - \delta\nu \ln{\frac{\omega_D}{\Delta_1 - \Delta}}.
\label{TcD}
\end{split}
 \end{gather}
Solving \eqref{TcD} gives Eq. \eqref{Tc} in the main body.

\section{Conductivity}
The integral \eqref{conduct} in the main body is computed  with complex analysis. 
Combination of Green functions 
\begin{gather}
    G^{A}(\ve,\bk)G^{R}(\ve,\bk) = \frac{1}{\left(\ve - E(\bk)\right)^2 + \big(\frac{1}{2\tau}\big)^2}
\end{gather}
can be substituted with the Dirac $\delta$-function $\delta\left(\ve - E(\bk) \right)$ in the limit $\frac{1}{\tau}\ll |\ve-E(\bk)|$:
\begin{gather}
    G^{A}(\ve,\bk)G^{R}(\ve,\bk)\approx \tau\delta\left(\ve-E(\bk)\right).
\end{gather}

 After integrating over $\ve$, Eq.\eqref{conduct} simplifies to

\begin{gather}
\label{conduct2}
\begin{split}
     \sigma_{ij}(T) = \frac{e^2}{2T} \sum_{\sigma} \int \frac{d \bk}{(2\pi)^2} \frac{\tau[E_{\sigma}(\bk)]}{4\cosh^2{\frac{E_{\sigma}(\bk,T)}{2T}}} 
       \frac{\partial E_{\sigma}(\bk)}{\partial k_i}\frac{\partial E_{\sigma}(\bk)}{\partial k_j}.
\end{split}
\end{gather}
From the last equation we obtain
the final Eqs. \ref{condTxx}-\ref{condTyy} presented in the main body.

\bibliographystyle{unsrt}
\bibliography{ref,bib}

\end{document}